# A brain-wide association study of DISC1 genetic variants reveals a relationship with the structure and functional connectivity of the precuneus in schizophrenia

Xiaohong Gong,[1,2*] Wenlian Lu,[1,3,4*], Keith M. Kendrick,[5*] Weidan Pu,[6,7*] Chu Wang,[2] Li Jin,[2] Guangmin Lu,[8,9] Zhening Liu,[6], Haihong Liu[10&], Jianfeng Feng,[1,3,4&]

[1] Centre for Computational Systems Biology, School of Mathematical Sciences, Fudan University, Shanghai, 200433, China, [2]State Key Laboratory of Genetic Engineering and MOE Key Laboratory of Contemporary Anthropology, School of Life Sciences, Fudan University, Shanghai, 200433, China, [3]Centre for Scientific Computing, University of Warwick, Coventry CV4 7AL, United Kingdom, [4]Fudan University-JinLing Hospital Computational Translational Medicine Centre, Fudan University, Shanghai, 200433, China, [5]Key Laboratory for Neuroinformation, School of Life Science and Technology, University of Electronic Science and Technology of China, Chengdu, 610054, China, [6]Medical Psychological Institute and [7]Institute of Mental Health, Second Xiangya Hopspital, Central South University, Changsha, 410011, China, [8]Department of Medical Imaging and [9]Jinling Hospital-Fudan University Computational Translational Medicine Center, Jinling Hospital, Nanjing University School of Medicine, Nanjing 210002, PR China, [10]Mental Health Center, Xiangya Hospital, Central South University, Changsha ,410008, China.

*These authors contributed equally to this work.

[&]Correspondence should be addressed to Prof. Jianfeng Feng, Centre for Computational Systems Biology, School of Mathematical Sciences, Fudan University, Handan Road 220, Shanghai, 200433, China. E-mail: jffeng@fudan.edu.cn, and Dr. Haihong Liu, Mental Health Center, Xiangya Hospital, Central South University, Changsha, 410008, China. E-mail: liuhaihong9@gmail.com

*Keywords:* Schizophrenia; DISC1; Precuneus; MRI; Voxel-wise association study; Link-wise association study

Abbreviated title: Association between DISC1 and MRI in precuneus

Number of pages: 35

Number of figures: 8

Number of Tables: 3

Number of multimedia: 0

Number of 3D models: 0

Number of words for Abstract: 233

**Conflict of interest**

The authors declare no competing financial interests.

**Acknowledgements**

This work was supported by the Royal Society Wolfson Research Merit Award, National Centre for Mathematics and Interdisciplinary Sciences (NCMIS) of the Chinese Academy of Sciences, the Key Program of National Natural Science Foundation of China (No. 91230201) to JFF, the Marie Curie International Incoming Fellowship from the European Commission (FP7-PEOPLE-2011-IIF-302421), and the New Century Excellent Talents in University of China (No. NCET-13-0139) to WLL, the National Natural Sciences Foundation of China (No. 61273309 to WLL, 30900404 to XHG and 91132720 to KMK, 81000587 to HHL), the National Basic Research Program of China (No. 2009CB522007) to XHG, and the Research Fund for Doctoral Program of Higher Education of China (No. 20100162120048) to HHL.

**Abstract**

The Disrupted in Schizophrenia Gene 1 (*DISC1*) plays a role in both neural signalling and development and is associated with schizophrenia, although its links to altered brain structure and function in this disorder are not fully established. Here we have used structural and functional MRI to investigate links with six *DISC1* single nucleotide polymorphisms (SNPs). We employed a brain-wide association analysis (BWAS) together with a Jacknife internal validation approach in 46 schizophrenia patients and 24 matched healthy control subjects. Results from structural MRI showed significant associations between all six *DISC1* variants and gray matter volume in the precuneus, post-central gyrus and middle cingulate gyrus. Associations with specific SNPs were found for rs2738880 in the left precuneus and right post-central gyrus, and rs1535530 in the right precuneus and middle cingulate gyrus. Using regions showing structural associations as seeds a resting-state functional connectivity analysis revealed significant associations between all 6 SNPS and connectivity between the right precuneus and inferior frontal gyrus. The connection between the right precuneus and inferior frontal gyrus was also specifically associated with rs821617. Importantly schizophrenia patients showed positive correlations between the six DISC-1 SNPs associated gray matter volume in the left precuneus and right post-central gyrus and negative symptom severity. No correlations with illness duration were found. Our results provide the first evidence suggesting a key role for structural and functional connectivity associations between *DISC1* polymorphisms and the precuneus in schizophrenia.

**Introduction**

Schizophrenia is a complex syndrome mainly defined by positive and negative symptoms of psychosis (Mueser and McGurk, 2004). It is increasingly viewed as a developmental disorder with onset of first major symptoms normally in late adolescence following a prodromal period lasting a number of years (Insel, 2010, Fusar-Poli et al., 2013). Current antipsychotic treatments mainly reduce positive psychotic symptoms but are less effective at treating the broad range of cognitive and emotional disturbances associated with the disorder, and may themselves also contribute to brain changes (Milev et al., 2005, MacDonald and Schulz, 2009, Lui et al., 2010, Ho et al., 2011).

Many neuroimaging and neurophysiological studies have reported structural and functional changes in the brains of schizophrenia patients, particularly dysconnectivity between frontal, parietal and temporal regions and the two brain hemispheres. These changes are also often associated with the symptom severity (Friston and Frith, 1995, Liang et al., 2006, Garrity et al., 2007, Zhou et al., 2007, Greicius, 2008, Lui et al., 2009, Huang et al., 2010, Lynall et al., 2010, Guo et al., 2012). Although there is considerable variability in reported findings, the default mode network (DMN) is often implicated (Meyer-Lindenberg et al., 2005, Bluhm et al., 2007, Garrity et al., 2007, Whitfield-Gabrieli et al., 2009).

A major focus has been on identifying key genetic contributions to schizophrenia, although robust links between specific neural circuitry changes and genetic polymorphisms are not reliably established. One gene frequently associated with schizophrenia is the Disrupted in Schizophrenia Gene 1 (*DISC1*), which has a protein scaffold function and influences both neuronal signalling and development (Brandon et al., 2009, Porteous et al., 2011, Thomson et al., 2013). A recent review of all genes associated with risk prediction for schizophrenia identified *DISC1* as the top candidate (Ayalew et al., 2012), although not all studies have found significant links (Brandon et al., 2009, Mathieson et al., 2012). Three common missense mutations in *DISC1* often associated with schizophrenia are R264Q (rs3738401), L607F (rs6675281) and S704C (rs821616) in Caucasian populations and S704C in the Chinese Han population (Qu et al., 2007, Brandon et al., 2009, Porteous et al., 2011). While *DISC1* variants, particularly L607F and S704C, are associated with altered development, structure or function in frontal and temporal cortical regions in both healthy subjects and schizophrenia patients (Duff et al., 2013, Thomson et al., 2013), findings are often inconsistent, do not incorporate internal validation steps and are usually seed-based rather than using an unbiased brain-wide association approach. Also, no studies have investigated altered resting-state functional connectivity associated with *DISC1*, although structural connectivity changes have been associated with S704C in healthy subjects (Li et al., 2013).

A semi-parametric regression model has been proposed to describe the association between SNPs in the same genetic pathway or nearby loci. In this way, the covariate effect of a single locus is modelled parametrically but the pathway interaction of multiple gene SNPs is modelled non-parametrically by using least-squares kernel machines (LSKM) (Liu et al., 2007, Kwee et al., 2008). This approach has been shown to be flexible for modelling high-dimensional interactions while allowing for covariates, and efficient for association mapping of quantitative traits (Ge et al., 2012).

In the current study we have used an imaging genetics approach, based on semi-parametric regression and LSKM, to investigate brain wide associations between six single nucleotide polymorphisms (SNPs) of the *DISC1* gene and grey matter (GM) and resting-state functional connectivity changes in the brains of both schizophrenia patients and healthy controls. Initial analysis was carried out on a combined group of controls and patients to increase statistical power. Additionally, we performed an internal validation of the robustness of our findings using a Jackknife approach, and investigated correlations between *DISC1* SNP-associations and symptom severity and illness duration in patients.

**Materials and Methods**

A schematic showing the overall experimental design used in the study is provided in Figure 1.

*Subjects.* A total of 46 schizophrenia patients and 24 healthy controls were recruited from the Second Xiangya Hospital, Central South University, China. The patients and healthy controls were right handed and matched for sex and education duration, although control subjects were slightly older (Table 1). All except one of the patients were Han Chinese. The patients were interviewed by trained psychiatrists and met the DSM-IV diagnostic criteria for schizophrenia. Patient symptoms were evaluated using the Positive and Negative Symptoms Scale (PANSS). Eight patients (17.4%) did not take any antipsychotics, and thirty-eight patients were receiving antipsychotics at the time of scan: thirty-seven patients (80.4%) were receiving second generation antipsychotics (SGAs) (clozapine, risperidone, quetiapine, olanzapine, sulpiride or aripiprazole), in which thirty-one patients (67.4%) received monotherapy and six patients（13%）received polytherapy with two SGAs, and only one patients (2.2%) were receiving combination therapy (combining a first-generation antipsychotics and an SGA). All medication doses were converted to chlorpromazine equivalence (50~1067 mg/day). The mean dosages of monotherapy and polytherapy with SGAs and combination therapy were 431.3 mg/day, 466.7mg/day and 400 mg/day, respectively. The mean duration of therapy for the patients was 32.2 weeks (range 2-144 weeks). The healthy controls were all assessed by structured interviews with experienced psychiatrists in accordance with DSM-IV criteria as being free of schizophrenia and other Axis I disorders. None had any neurological diseases or suffered from clinically significant head trauma or had a history of any substance dependence. Written informed consent was obtained from all individual participants, and research procedures and ethical guidelines were followed in accordance with the Institutional Review Board (IRB) of the Second Xiangya Hospital, Central South University. Genomic DNA was extracted from whole blood using standard protocols.

*Genotyping of DISC1 variants.* The reference sequence of the *DISC1* gene was acquired from the UCSC Genome Browser (NM_018662). Three segments containing functional variants rs3738401 (R264Q), rs6675281 (L607F) and rs821616 (S704C) were sequenced in all samples using BigDye Terminator version 3.1 in ABI 3100 sequencers (Applied Biosystems, Foster City, CA). The forward primer of rs3738401 (R264Q) was 5'- GTT CCT TTC CCC AGC AGT G -3' and the reverse was 5'-AGA ATG CAT GTC ACG CTC T -3'. The forward primer of rs6675281 (L607F) was 5'-GAT GGC TTC ACC AAT GGA AC -3' and the reverse was 5'-CAG GTT GAG ACA GGG AA AGA -3'. The forward primer of rs821616 (S704C) was 5'- TGT CTC AGC TGC AAG TGT CC -3' and the reverse was 5'- ATG CCA AAA GTT GGG TTT TT -3'.

After sequencing, ten *DISC1* variants were identified, six of which were common SNPs with minor allele frequency > 5% in our samples. These six SNPs: rs3738401 (R264Q), rs2738880, rs12133766, rs1535530, rs821616 (S704C) and rs821617, were included in the following analysis while four rare variants: rs56020408, rs6672782, rs11122391, rs6675281 (L607F) were excluded due to their low frequencies in our samples.

Linkage disequilibrium (LD) analysis of SNPs was tested using Haploview 4.2 software. D' and $r^2$ for each pair of SNPs were calculated in LD analysis. No pairwise SNPs showed a high level of LD except the pair of rs821616 and rs821617 (D'=1.0, $r^2$=0.69).

*Structural MRI acquisition and pre-processing.* All image data were acquired using a 1.5T Siemens MRI scanner. High-resolution whole brain volume T1-weighted images were acquired sagittally with a 3D spoiled gradient echo(SPGR) pulse sequence (repetition time,

12.1ms; echo time, 4.2ms; flip angle,15 degree; field of view = 240×240 mm$^2$; acquisition matrix, 256×256; thickness, 1.8 mm; number of excitations, 2; 172 slices.)

All T1-weighted structural data were pre-processed with SPM5 software package (http://www.fil.ion.ucl.ac.uk/spm) based on General Linear Model and Gaussian Random Field theory (Friston et al., 1995). The normalization, segmentation, and modulation were completed in one step, resulting in modulated GM. In the normalization, all images were spatially normalised to the T1-weighted template in the Montreal Neurological Institute (MNI) space, and were re-sampled into a final voxel size of mm$^3$. The modulated images were then smoothed with a full-width at half-maximum (FWHM) 8-mm Gaussian kernel for further analysis. Identification of brain regions was performed used the automated anatomical labelling (AAL) atlas which parcellates the brain into 90 regions of interest (ROIs; 45 in each hemisphere) (Tzourio-Mazoyer et al., 2002). In the following, we only use Talairach coordinates to locate the voxels in brain, which is a common coordinate system used in fMRI and transcranial stimulation studies of brain regions (Bankman, 2008). An alternative system is the Montreal Neurological Institute and Hospital (MNI) coordinate, which can be transformed to the Talairach coordinate by the following equations:

$$x = 0.99 \times X, y = 0.9688 \times Y + 0.046 \times Z, z = -0.0485 \times Y + 0.9189 \times Z \quad if \ Z \geq 0$$

$$x = 0.99 \times X, y = 0.9688 \times Y + 0.042 \times Z, z = -0.0485 \times Y + 0.839 \times Z \quad if \ Z < 0$$

where [X, Y, Z] is the MNI coordinate and [x,y,z] is the responding Talairach coordinate.

*Functional MRI acquisition and pre-processing.* For fMRI, a total of 180 volumes of Echo Planar Imagining (EPI) images were obtained axially (repetition time, 2000 ms; echo time, 40 ms; slices, 20; thickness, 5mm; gap, 1mm; field of view (FOV), 24×24 mm$^2$; resolution, 64× 64; flip angle, 90 °). Prior to pre-processing, the first 10 volumes were discarded for scanner stabilisation and the subjects' adaptation to the environment.

Pre-processing of fMRI data was then conducted using SPM5 and a Data Processing Assistant for Resting-State fMRI (DPARSF) (Chao-Gan and Yu-Feng, 2010). In all cases head movements did not exceed the criterion of greater than ±1.5mm or ±1.5°. The functional scans were firstly corrected for within-scan acquisition time differences between slices, and then realigned to the middle volume to correct for inter-scan head motions. Subsequently, the functional scans were spatially normalised to a standard template (MNI space) and resampled to 3 mm$^3$. After normalisation the BOLD signal of each voxel was firstly detrended to abandon linear trend and then passed through a band-pass filter (0.01-0.08 Hz) to reduce low-frequency drift and high-frequency physiological noise. Finally, nuisance covariates including head motion parameters, global mean signals, white matter signals and cerebrospinal signals were regressed out from the BOLD signals.

The brain regions of interest were allocated by a special approach that will be addressed in subsection of the link-wise association study (LWAS) below. To locate their brain regions of anatomy, the automated anatomical labelling (AAL) atlas were still used as mentioned above (Tzourio-Mazoyer et al., 2002).

*Multi-locus approach and least square kernel machines.* Let N be the number of unrelated subjects. For each subject i, let $Y_i(v)$ denote the quantitative imaging trait (for example, gray matter volume in VBM or correlation coefficient) at a particular voxel or a link v. $X_i$ is a $q \times 1$ vector of the (non-SNP) covariates. In this study, age, sex and an intercept are included as covariates (here q=3), which comprise $X_i$. Let $G_i = [G_{i,1}, \cdots, G_{i,S}]^T$ be the $S \times 1$ vector with the element $G_{i,s}$ being the genotype for the SNPs of subject i, which is coded to be the number of copies of the minor allele that subject i possesses for SNPs, and takes the values of 0, 1, or 2. The semi-parametric model for a given voxel is:

$$Y_i(v) = X_i^T \beta_K(v) + h(G_i) + \varepsilon_{i,K}(v), \qquad i = 1, \cdots, N$$

for all voxels v. Here, $\beta_K(v)$ is the $q \times 1$ regression parameter vector and the errors $\varepsilon_{i,K}(v)$ are assumed to be normally distributed with mean 0 and standard deviation $\sigma_K$. $h(\cdot)$ denotes the non-parametric function of the SNPs, defined in a function space $\mathcal{H}_K$, of which the kernel matrix ($N \times N$) is positive definite and depends on the genotype data (Liu et al., 2007). In particular, following the following setup (Kwee et al., 2008, Ge et al., 2012), the kernel function (matrix) is defined:

$$k(G_j, G_k) = \frac{1}{2S} \sum_{s=1}^{S} IBS(G_{j,s}, G_{k,s})$$

where $\sum_{s=1}^{S} IBS(G_{j,s}, G_{k,s})$ denotes the number of alleles shared IBS by subjects j and k at the SNPs, and takes values 0, 1, or 2. Here, we assume $IBS = -1$ if one individual has missing genotype and emphasise that it does not affect the results if picking other values. The healthy control and schizophrenia subjects were mixed into this model to increase the statistic power, which is a routine in imaging genetic association study towards illness (Hibar et al., 2011a, Hibar et al., 2011b, Ge et al., 2012), see (Hibar et al., 2011a) for a review and (Hibar et al., 2011b, Ge et al., 2012, Vounou et al., 2012) for examples.

By using a connection to Linear Mixed Models, a score statistic based on the null (non-SNP) model can be used to test the effect of multiple SNPs on the traits (Liu et al., 2007):

$$Q_\tau = \frac{1}{2\hat{\sigma}_K^2}(Y - X\hat{\beta})^T K (Y - X\hat{\beta})$$

where $Y = [Y_1, \cdots, Y_N]^T$ and $X = [X_1, \cdots, X_N]^T$, $\hat{\sigma}_K$ and $\hat{\beta}$ are the estimators of the linear model $Y = X\beta + \varepsilon$. The Satterthwaite method is then used to approximate the distribution of Q by a scaled $\chi^2$-distribution, $\kappa \chi_v^2$, where the scale parameter $\kappa$ and the degrees of freedom $v$ were calculated in (Liu et al., 2007) as:

$$\kappa = \frac{\tilde{I}_{\tau\tau}}{2\tilde{e}}, \quad v = \max\left\{1, \frac{2\tilde{e}^2}{\tilde{I}_{\tau\tau}}\right\}$$

where $\tilde{I}_{\tau\tau} = I_{\tau\tau} - I_{\tau\sigma^2} I_{\sigma^2\sigma^2}^{-1} I_{\tau\sigma^2}^T$, $I_{\tau\tau} = tr(P_0 K)^2/2$, $I_{\tau\sigma^2} = tr(P_0 K P_0)/2$, $I_{\sigma^2\sigma^2} = tr(P_0^2)/2$, and $P_0 = I - X(X^T X)^{-1} X^T$. The significance of the test can then be assessed by comparing the scaled score statistic to the chi-squared distribution with $v$ degrees of freedom.

In addition, when considering LWAS, v stands for the index of functional link between a pair of ROI and $Y_i(v)$ stands for the (group) correlation coefficient at the link v.

*Group correlation coefficients between ROIs.* After data pre-processing, the fMRI data were extracted into voxel-wise time courses. Considering two ROIs, each of which have a number of time courses, denoted by $A = \{x_1(t), \cdots, x_n(t)\}$ and $B = \{y_1(t), \cdots, y_m(t)\}$, $t = 1, 2, \cdots, T$, n and m are the number of voxels in these two ROIs respectively, to calculate the group correlation coefficients, the spatial Principle Component Analysis (PCA) was used to extract the principle time course of each ROI. Let $PC_A(t) = \sum_{i=1}^{n} \alpha_i x_i(t)$ be the first principle component of A and $PC_B(t) = \sum_{j=1}^{m} \beta_j y_j(t)$ be the first principle component of B. Thus, the group correlation coefficient between the ROIs A and B is defined as the Pearson correlation coefficient between $PC_A(t)$ and $PC_B(t)$:

$$CC_{A,B} = \frac{\text{cov}(PC_A, PC_B)}{\sqrt{\text{var}(PC_A)\text{var}(PC_B)}}$$

*Voxel-wise association analysis (VWAS).* The T1-weighted structural data were analysed for the association with *DISC1* variants using the multi-locus approach of a semi-parametric regression model (see above). Six SNPs of *DISC1* were considered in the model, and covariate effects such as sex and age modelled parametrically (i.e. linearly), including an intercept column with all components equal to 1. The interaction of SNPs was modelled non-parametrically using a least-squares kernel machines (LSKM) approach which allowed a flexible function of the joint effect of multiple SNPs on the imaging traits by specifying a kernel function (Kwee et al., 2008). The SPM5 based General Linear Model was used to estimate regression parameters and achieve residual error vectors. By using a connection to Linear Mixed Models, a score statistic based on the null (no-SNP) model was used to test the effect of multiple SNPs on traits followed by a Satterthwaite approximation test (Liu et al., 2007).

Thus, chi-square statistics at all voxels form a statistical parametric mapping. Two analyses are performed in this study: highest peak identification and largest cluster localisation. A region-wise Bonferroni correction (i.e. $\times 45$) was performed to obtain a corrected p-value (p<0.05 considered significant). Peak identification is achieved by searching for the voxel in the statistical parametric mapping with the largest value (smallest p-value). For cluster localization, each cluster is formed as a set of contiguous voxels with chi-square values exceeding a pre-defined cluster forming threshold, where contiguity is defined by an order-18 neighbourhood (voxels need at least a common edge to be connected). The cluster-forming threshold is set to an uncorrected p-value threshold of 0.02. The largest cluster size is defined in RESELs (Resolution Elements) by the random field theory, number of voxels, as well as their GM volumes (Ge et al., 2012). The corresponding p-value and its region-wise Bonferroni correction of each cluster is calculated using "stat_thresh" in the NS toolbox of SPM5.

Peak and cluster analysis was also carried out for individual SNP, where p-values survived both region-wise and SNP-wise Bonferroni correction, i.e. $\times(45\times 6)$.

Each SNP that survived in the SNP-wise VWAS above was picked up to divide the whole sample into two groups: one group comprising the subjects for whom the major allele of this SNP occurs twice, i.e. whose number of copies of the major allele is 2, and the remaining, i.e. whose number of copies of the major allele is 0 or 1. Then, student t-tests were conducted to compare the GM volumes of the significant voxels in each brain region (AAL) surviving in the SNP-wise VWAS for this SNP in the two groups.

*Link-wise association analysis (LWAS).* The functional links associated with *DISC1* SNPs were analysed by LWAS using regions with significant voxels identified by the previous VWAS analysis as seeds. Here, a 5mm radius sphere centred at the significant peak voxel in each AAL brain region identified by VWAS, acted as the seed ROI and the whole brain was parcellated into 1072, non-overlapping ROI cubes with a side-length of 12 mm. We then obtained the functional correlation between the seed and cube ROIs by calculating the group correlation coefficients (GCCs) of the time courses from fMRI data (see above).

Analogous to VWAS, we utilised the multi-locus approach of a semi-parametric regression model (see above) to relate the functional link from the seed ROI to the others with SNPs. The six SNPs were considered in the model and the LSKM approach was used to achieve statistical parametric mapping of chi-square test statistics. An ROI-wise Bonferroni correction was performed by multiplying uncorrected p-values by the number of the cube ROIs, i.e. ×1072.

We also studied the correlation between each single SNP and the functional connection from the seed ROI to each non-overlapping cube ROI. The Bonferroni correction used both ROI-wise and SNP-wise, i.e. ×(1072×6). Thus, in this model we let v be the index of cube ROI and $Y_i(v)$ denote the group correlation coefficients from the cube ROI v to the predefined seed. Here we report correlation coefficients without Fisher r-z transformation, however we have confirmed that the same overall results are obtained from our LWAS analysis following a Fisher transformation (data not shown).

Each SNP that survived in the SNP-wise LWAS above was picked up to divide all of the subjects into two groups: one group comprising subjects whose number of copies of the minor allele is 2, and the other of those whose number of copies of the minor allele is 0 or 1. A, students t-test was then conducted to compare the correlation coefficients of the significant links surviving in the SNP-wise for this SNP-wise LWAS in the two groups.

*Correlations with PANSS scores, illness durations and medication dose*. Pearson correlations were used to investigate associations between PANSS scores, illness durations and medication dose (daily dose in chlorpromazine equivalents), and GM volumes (VBM) respectively that survived the VWAS analysis, and the strength of functional links that survived the LWAS analysis.

*Internal validation by $d_n$-Jackknife.* Jackknife, like bootstrap, is a widely-used technique for resampling to verify the accuracy and robustness of a statistical approach (Miller, 1974). In addition to delete-one resampling, the general $d_n$-Jackknife approach (i.e. delete-(n-$d_n$) Jackknife) approximates the *true* distribution of the statistics of interest by selecting all sample subsets of the size $d_n$ *without replacement*, and is particularly useful for cases with small sample sizes. The empirical distribution by $d_n$-Jackknife converges to the true one asymptotically, and especially satisfies first and second-order properties (Babu and Singh, 1985, Bertail, 1997). Here we utilised Jackknife using two steps. (1). Resampling: we selected 90% (63 subjects) of samples without replacement from the original whole sample set (71 subjects); (2). Re-calculating: we calculated the chi-square statistics and their uncorrected p-values by the same multi-locus model and LSKM approach (see above) based on this sub-sample set. Since we cannot carry out calculations for all sub-sample sets without replacement (i.e. $>10^9$), we randomly selected the subsample set 10000 times with equal probabilities to generate the empirical distribution. This proved to be robust despite the random selection of 10000 sub-sample sets (data not shown). Instead of the statistic of interest (chi-square), we show the empirical distribution of the uncorrected p-values for the chi-square statistics and compare it with the original (for all samples).

We did not use bootstrap by resampling *with replacement* since in this case the replacement causes unexpected replication of subjects and increases the number of zero elements in the kernel matrix of LSKM. This might influence the order of the chi-square statistics and thus make their values incomparable for different resampling processes.

**Results**

**VWAS analysis**

The VWAS analysis based on all six SNPs found a number of voxels that survived after region-wise Bonferroni correction with four peak voxels, significantly associated with *DISC1* variants. Using Talairach co-ordinates they were located in the right precuneus ([14,-52,36], corrected p-value of 0.0076), right middle cingulate gyrus ([14,-50,34], corrected p-value of 0.0087), right post-central gyrus ([38,-40,64], corrected p-value of 0.0074), and left precuneus ([-10,-64,64], corrected p-value of 0.025), respectively (Table 2, Figures 2 and 3). The most significant peak voxel in the right precuneus was not isolated but also belonged to the largest significant cluster. All the voxels in this cluster were in the right precuneus and neighbouring right middle cingulate gyrus. Two individual SNPs, rs2738880 and rs1535530, showed significant associations with the voxels located in the same regions. Voxels in the right post-central gyrus were associated with rs2738880 (corrected p-value of 0.0279) and there was a trend for the left precuneus as well (corrected p-value of 0.063). Voxels in the right precuneus and right middle cingulate gyrus were associated with rs1535530 (corrected p-values of 0.0165 and 0.0223, respectively; Table 2, Figures 2 and 3). For the cluster analyses, the only significant cluster was also found in right precuneus and right middle cingulate gyrus and this was associated with rs1535530. Carriers of the TT genotpye in rs1535530 showed increased GM volumes for these significant voxels in both right precuneus and right middle cingulate gyrus compared to the individuals with CC/CT genotypes (Figure 4a-b) but genotypes of rs2738880 were not significantly correlated with GM volumes for these significant voxels in the post-central gyrus (Figure 4c).

**LWAS analysis**

A 5mm radius ROI sphere centred at the peak voxel in the right precuneus identified with the six common SNPs was set as a seed and its functional connectivity with the remaining cube ROIs (1072 ROIs) was investigated. The LWAS analysis identified a significant cube ROI with a region-wise Bonferroni corrected p-value 0.0154 in the right triangular inferior frontal gyrus, as shown in Figure 5 and Table 3. This significant cube ROI was also associated with SNP rs821617 alone (p = 0.0346 after correction) (Figure 5, Table 3). Here genotypes of this SNP were significantly correlated with the correlation coefficients of this link (Figure 4d).

In the same way, we also set 5mm radius ROI spheres centred at the peak voxel in the left precuneus, the right post-central gyrus and right middle cingulate gyrus, identified with the six common SNPs, as seeds, and investigated functional connectivity with the remaining cube ROIs. However, no significant cube ROIs had significantly associated functional connections (Table 3).

**Links with symptom severity and illness duration as well as medication dose in schizophrenia patients**

After the VWAS and LWAS, we calculated Pearson correlations between the SNP-associated GM volumes and functional connections and PANSS scores as well as illness durations and medication dose in the schizophrenia patients (Tables 2 and 3, Figure 3). The GM volume of voxels in the left precuneus associated with all six *DISC1* SNPs was significantly positively correlated with negative, but not positive or general psychopathology PANSS scores (Table 2, Figure 3). The only other region showing significant GM volume correlations with PANSS scores was the right post-central gyrus which was also significantly positively correlated with negative but not positive symptoms. There was also a trend towards a significant positive correlation with general scores (Table 2). For individual SNPs, the voxels in the right post-central gyrus associated with rs2738880 were significantly positively correlated with negative but not positive scores and there was a trend towards positive correlation with general scores (Table 2). We did not find significant correlations between PANSS symptom scores and voxels in the right precuneus and middle cingulate gyrus associated with either all SNPs or the single SNP rs1535530.

There were no significant correlations between DISC1-associated GM volumes or functional connections and illness duration (p>0.5 and N=46 in all cases: GM – right middle cingulate gyrus – r=-0.037, p = 0.808; right postcentral gyrus – r = 0.045, p = 0.769; right precuneus – r = -0.045, p = 0.768; left precuneus – r = 0.081, p = 0.593: Functional connections – right precuneus to right inferior frontal gyrus – r = -0.102, p = 0.5;). There were also no significant correlations with medication dose (p>0.1 and N=38 in all cases: GM – right middle cingulate gyrus – r=-0.13, p = 0.50; right postcentral gyrus – r = -0.25, p = 0.17; right precuneus – r = 0.029, p = 0.88; left precuneus – r = -0.021, p = 0.91: Functional connections – right precuneus to right inferior frontal gyrus – r = -0.17, p = 0.36;).

**Internal validation of VWAS and LWAS**

For VWAS, first we selected the four peak voxels located in the right precuneus ([14,-52,36] in Talairach space coordinate), right middle cingulate gyrus ([14,-50,34]), right post-central gyrus ([38,-40,64]), and left precuneus ([-10,-64,64]), respectively (Table 2). Their empirical distributions by $d_n$-Jackknife indicated that the original chi-square statistics shown by its un-corrected p-values together with the threshold of the brain region-wise Bonferroni correction (0.05/45) stayed within the 5%-95% confidence interval (Figures 7a-d). Next we illustrated the robustness of our results by counting: (1) The distribution of each of the 90 AAL brain regions containing a GM peak voxel with respect to the uncorrected p-values via 10000 times of random resampling; (2) The percentage of times that each brain region contained a significant voxel after brain region-wise (AAL) Bonferroni correction, i.e. uncorrected p-values $< 0.05/45$, across the 10000 times of random resampling. The four brain regions identified (the right and left precuneus, right middle cingulate gyrus and right post-central gyrus) contribute more than 95% of the peak voxels, with the precuneus (left and right) contributing more than 40% (Figure 7e). The same regions also contained a significant percentage of voxels across the 10000 times of resampling: the right precuneus had $> 65\%$, and the left precuneus, the right middle cingulate gyrus and post-central gyrus $> 70\%$, (Figure 8f).

To validate the LWAS results we employed similar approaches using functional connections, i.e. the Pearson correlation coefficients between the seed ROI and the non-overlapping cube ROIs (see *Methods*). First, we validated the uncorrected p-values of the chi-square statistics of the significant link between ROIs in the right precuneus (seed) and right triangular inferior frontal gyrus (cube ROI). The empirical distribution of uncorrected p-value (equivalent to chi-square statistics) contain the original ones and the threshold (cube ROI-wise Bonferroni correction, i.e. 0.05/1072) in 5%-95% confidence intervals. Here the link between the right precuneus and right triangular inferior frontal gyrus has a Gaussian-like distribution for the logarithmic p-values (See Figures 8a). For the robustness of the LWAS results we showed that with the seed ROI in the right precuneus, the cube ROI located in the right triangular inferior frontal gyrus contributed more than 70% (73.5%) of the peak links among all 1072 cube ROIs and was significant near 50% of the 10000 times of resampling (Figures 8bc).

These self-validation results showed the robustness and accuracy of the statistical approaches. All p-values of the peak voxels of the significant (AAL) ROIs and the significant functional link obtained from analysing the whole dataset are located in the 5%-95% confidence interval of the distributions by Jackknife bootstrap. That is to say, there is no evidence of heterogeneity in the data that can reject the original p-values. The p-value thresholds (0.05/45 for VWAS and 0.05/1072 for LWAS), the red arrows in Figures 7a-d and Figure 8a, are all located before (the right precuneus, the right post-central gyrus and the right middle cingulate gyrus) or near (the left precuneus and the link between the right precuneus and triangular inferior frontal gyrus) 50% in the histograms. In addition, these ROIs and the link contributed most to the partial of the peak and significant voxel/link over the whole brain gray matter or cube ROIs. Both indicate strong support for the original p-values obtained in the distribution of significant effects.

**Discussion**

Overall our results from a brain-wide association analysis provide evidence that the precuneus is the main brain region showing significant links with *DISC1* polymorphisms and schizophrenia both in terms of GM volumes and functional links. The right post-central gyrus was the only other region showing this relationship with schizophrenia for GM volume. Overall in healthy subjects and schizophrenia patients, rs2738880 showed significant links with the right post-central gyrus GM volume and a trend towards significance for the link within the left precuneus(peak voxels), as well as the right and left precuneus and middle cingulate gyrus GM volumes (clusters). Additionally rs1535530 was associated with GM volume (both peak voxels and clusters) in the right precuneus and middle cingulate gyrus.

Functional connectivity between the right precuneus and the right triangular inferior frontal gyrus was associated with rs821617. However, only GM volumes in the left precuneus and right post-central gyrus were positively correlated with negative symptom severity in patient. Thus *DISC1* polymorphisms may be important in regulating precuneus and post-central gyrus function and contribute to negative symptom severity in schizophrenia.

In support of our findings a previous study has provided some preliminary evidence using an independent component analysis for an association between precuneus activation during an auditory oddball task and the DISC1 SNP rs 821616 (S704C) in schizophrenia (Liu et al., 2009). The precuneus is a key component in the brain default circuit and as a "rich club" member has many long distance connections and able to exert widespread influence on both cortical and limbic functions (Fransson and Marrelec, 2008, van den Heuvel et al., 2012). There is increasing evidence for a key involvement of the precuneus in schizophrenia. We have previous reported that altered resting state functional connectivity in a parietal circuit including the precuneus was the most altered in schizophrenia patients from Taiwan (Guo et al., 2012). The precuneus also shows altered resting-state activity (Kühn and Gallinat, 2013) and task-related deactivation in schizophrenia (Garrity et al., 2007, Whitfield-Gabrieli et al., 2009). Some studies have also reported precuneus GM volume reductions (Theberge et al., 2007, Morgan et al., 2010, Tanskanen et al., 2010), although others have not (Glahn et al., 2008, Guo et al., 2012). The precuneus is involved in many different behavioural functions (Cavanna and Trimble, 2006) including reflective and self-related processing (Kjaer et al., 2002, Lou et al., 2004), awareness and conscious information processing (Kjaer et al., 2001, Vogt and Laureys, 2005, Cavanna, 2007), empathy (Harvey et al., 2013), episodic memory (Lundstrom et al., 2003, Lundstrom et al., 2005, Dorfel et al., 2009) and visuo-spatial processing (Wenderoth et al., 2005). Many of these functions are impaired in schizophrenia, and most notably studies have reported altered precuneus function associated with impaired self-processing (Zhao et al., 2013), insight (Faget-Agius et al., 2013) and empathy (Harvey et al., 2013) in schizophrenia patients.

A significant positive correlation was found between *DISC1* associated GM volume in the left precuneus and negative PANSS scores. Several previous studies have reported links between *DISC1* polymorphisms and positive symptoms associated with structure and function of the hippocampus and prefrontal cortex (Callicott et al., 2005, Di Giorgio et al., 2008, Szeszko et al., 2008). Since schizophrenia is associated with extensive overall GM loss in the brain it is perhaps surprising that we observed a positive correlation between negative symptom severity and GM volume (Theberge et al., 2007, Glahn et al., 2008, Lui et al., 2009, Guo et al., 2012). However, similar positive correlations with negative symptoms have been reported previously (Nesvag et al., 2009). A potential explanation may be that increased GM volume in the *DISC1* associated region of the precuneus reflects compensatory changes resulting in increasing dysfunction.

In contrast with the precuneus, many studies have reported reduced GM volume in the post-central gyrus in schizophrenia (Glahn et al., 2008, Tanskanen et al., 2010, Guo et al., 2012). The post-central gyrus is engaged in somatosensory and motor processing and both are impaired in schizophrenia (Dazzan and Murray, 2002). The post-central gyrus is also involved in processing of emotional faces and shows altered responses to fear faces in Caucasian and Chinese schizophrenia patients (Phillips et al., 1999, Li et al., 2012). Further, a recent study has reported an association between the *DISC1* variant rs1538979 and post-central gyrus activation in schizophrenia patients in the Hayling sentence completion task (Chakirova et al., 2011). Consistent with these broad functions of the post-central gyrus we found that its GM volume was, associated with *DISC1* variant rs2738880 and positively correlated with negative PANSS scores.

The *DISC1* protein promotes growth of dendritic spines and functions presynaptically at glutamatergic synapses (Brandon et al., 2009, Porteous et al., 2011, Maher and LoTurco, 2012). In the parietal cortex *DISC1* is localized on dendritic spines, which are primarily glutamatergic, and there is evidence for pathology of cortical dendritic spines in schizophrenia (Kirkpatrick et al., 2006, Bennett, 2011). Indeed, it has been hypothesized that the cognitive and negative symptoms of schizophrenia are produced by hypofunction of cortical glutamatergic transmission (Marsman et al., 2013). The precuneus also shows increased activity following treatment NMDA receptor antagonists such as ketamine and memantine (Deakin et al., 2008, Lorenzi et al., 2011). Thus, *DISC1* associated increases in precuneus GM volume may reflect compensation for reduced glutamatergic signalling leading to increased dysfunction and negative symptom severity.

The functional link between the precuneus and IFG is associated with episodic and working memory, which are both disrupted in schizophrenia and these regions show reduced activation during decision-making in an item recognition task in patients (Paulus et al., 2002, Lundstrom et al., 2005, Kim et al., 2009, Grillon et al., 2010). Such cognitive dysfunctions are not strongly reflected in PANSS scores, which may explain the absence of correlations with this functional link. Working memory impairments in schizophrenia have also been linked with genetic susceptibility and *DISC1* polymorphisms with working memory performance (Friston et al., 1995, Park et al., 1995).

Of the individual *DISC1* polymorphisms linked with GM volumes and functional connectivity, rs2738880 and 1535530 are located on introns, and rs821617 in an exon leading to the change of amino acid (K800R) in *DISC1* protein isoform b (NM001164538; http://www.ncbi.nlm.nih.gov/protein/NP_001158010.1). To date none of these polymorphisms have reported associations with schizophrenia and could be causative variants affecting *DISC1* expression, or closely linked with others that are. Although

rs2738880 and 1535530 are close to the missense mutation L607F, linkage disequilibrium analysis revealed only a low linkage (D'=0.872, $r^2$=0.16). On the other hand rs821617 has a tight linkage with rs821616 (D'=1.0, $r^2$=0.69) which has been associated previously with structural changes in frontal and temporal but not parietal regions (Duff et al., 2013). While the current study had insufficient subjects to establish associations with different *DISC1* allele carriers, rs1535330 and rs821617 CC/CT and GG/AG carriers were significantly different from TT and AA ones for GM volumes and functional connectivity respectively. There was also a trend for rs2738880 GG/GA carriers to have higher genenal PANSS scores (See Figure 6).

The major allele frequency of rs6675281(L607F) in European and African populations reported by 1000 Genomes project and HapMap is between 0.805 and 0.842, while it is 1.0 in Asian populations, suggesting a significant genetic background difference in these populations. Consistent with these data, we found that this locus is a low-frequency polymorphism in Han Chinese population with the major allele frequency of 0.98. Although this SNP has frequently been reported to be associated with schizophrenia in Caucasian population, this could not be replicated in Chinese population due to the absence or extreme rarity of the minor allele. Similar to rs6675281, a difference of major allele frequency of another SNP rs821616 (S704C) exists between Caucasian (0.300) and Asian (0.068) populations, suggesting this variant may play a different role in susceptibility to schizophrenia in different populations.

Given that a number of previous studies both in psychiatric patients and transgenic mouse models have emphasised links between DISC1 and structural and functional changes in the frontal cortex and hippocampus (Callicott et al., 2005, Duff et al., 2013, Thomson et al., 2013), it is perhaps surprising that our brain-wide association study failed to support this, other than in terms of the functional link between the precuneus and the triangular inferior frontal gyrus. This may reflect the fact that we only found links with novel DISC1 polymorphisms, whereas previous studies showing DISC1 associations with the frontal cortex and hippocampus have focussed mainly on L607F (rs6675281) and S704C (rs821616). In our study we did not find any associations with either of these two polymorphisms although, as discussed above, rs2728880 and 1525530 are close to L607F and rs821617 has a tight linkage with S704C.

A limitation of this study is that there were insufficient numbers of unmedicated schizophrenia patients (8/46) to assess if neuroleptic medications per se contributed to our findings. However, there were no correlations between structural or functional connectivity associations with *DISC1* polymorphisms and medication dose in patients, which suggests that antipsychotic drugs were unlikely to have had a significant influence on our findings.

In summary we have shown that both GM volume and functional connectivity of the precuneus are associated with *DISC1* variants and negative (GM) symptom severity. Additionally, right post-central gyrus GM volume is also associated with negative symptoms and rs2738880. The *DISC1* related GM volume changes were all positively correlated with negative symptoms suggesting that compensatory increases in volume in these regions may contribute to dysfunction.

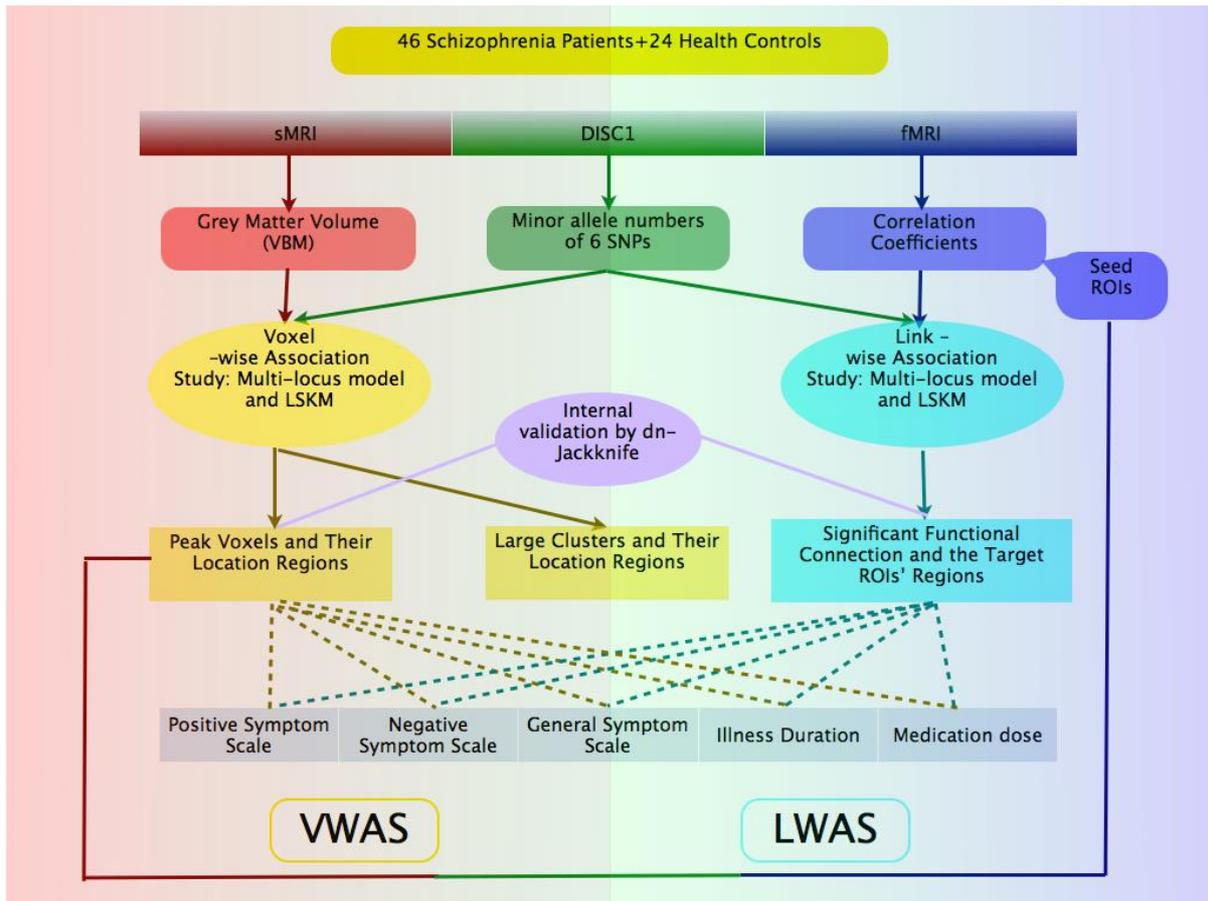

**Figure 1.** Flow chart showing approaches and experimental designs.

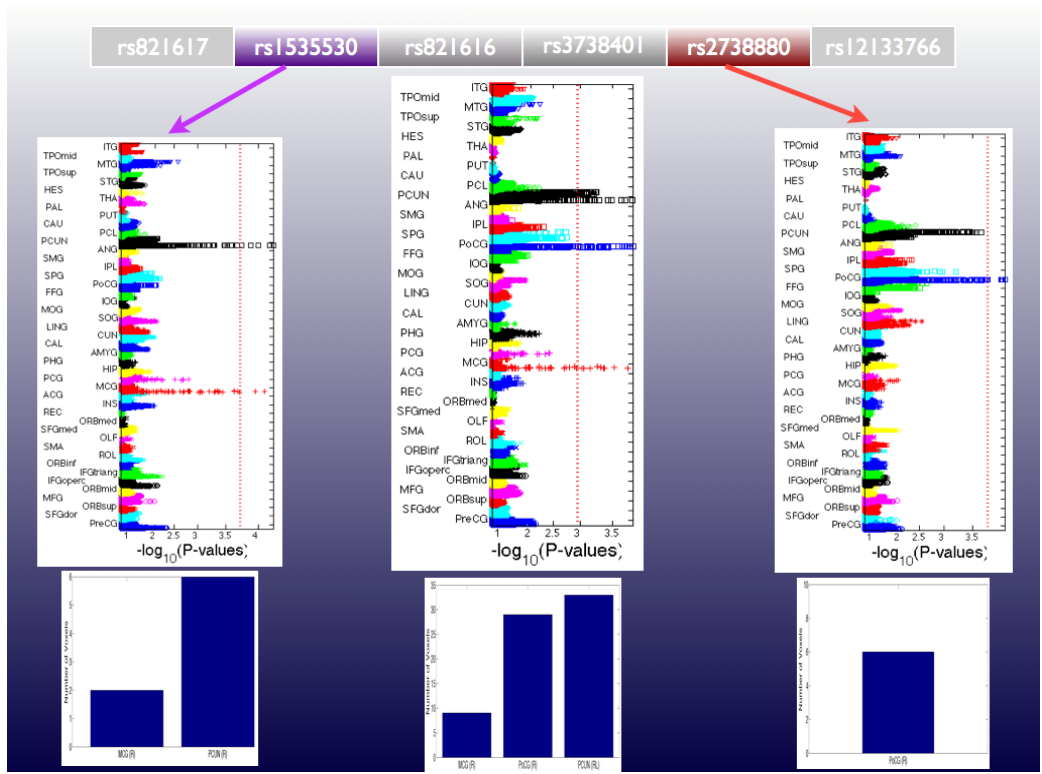

**Figure 2.** Manhattan plots of the voxel-wise distribution of uncorrected p-values by VWAS with all six SNPs in DISC1 (middle), rs1535530 (left) and rs2738880 (right) with respect to AAL brain regions. The histograms show corresponding numbers of significant voxels from the peak analysis for the significant regions. PoCG − post-central gyrus, PCUN − precuneus, MCG − middle cingulate gyrus, (L) − left, (R) − right. Abbreviations: AMYG − amygdala; ANG − angular gyrus; ACG − anterior cingulate gyrus; CAL − calcarine cortex; CAU − caudate; CUN − cuneus; FFG − fusiform gyrus; HER − Herschl's gyrus; HIPP − hippocampus; IOG − inferior occipital gyrus; IFGoper − inferior frontal gyrus opercular; IFGtriang − inferior frontal gyrus triangular; IPL − inferior parietal lobule; ITG − inferior temporal gyrus; INS − insula; LING − lingual gyrus; MCG - middle cingulate gyrus; MFG − middle frontal gyrus; MOG − middle occipital gyrus; OLF − olfactory; ORBinf − inferior orbitofrontal cortex; ORBmed − medial orbitofrontal cortex; PAL − pallidum; PCL − paracentral lobule; PHG − parahippocampal gyrus; PoCG − postcentral gyrus; PCG − posterior cingulate gyrus; PreCG − precentral gyrus; PCUN −

precuneus; PUT – putamen; REC – rectus gyrus; ROL – Rolandic operculum; SOG – superior ocular gyrus; SFGdor – dorsal superior frontal gyrus; SFGmed – medial superior frontal gyrus; SPG – superior parietal gyrus; SMA – supplementary motor area; SMG – supramarginal gyrus; TPOmid – middle temporal pole; TPOsup – superior temporal pole; THA – thalamus.

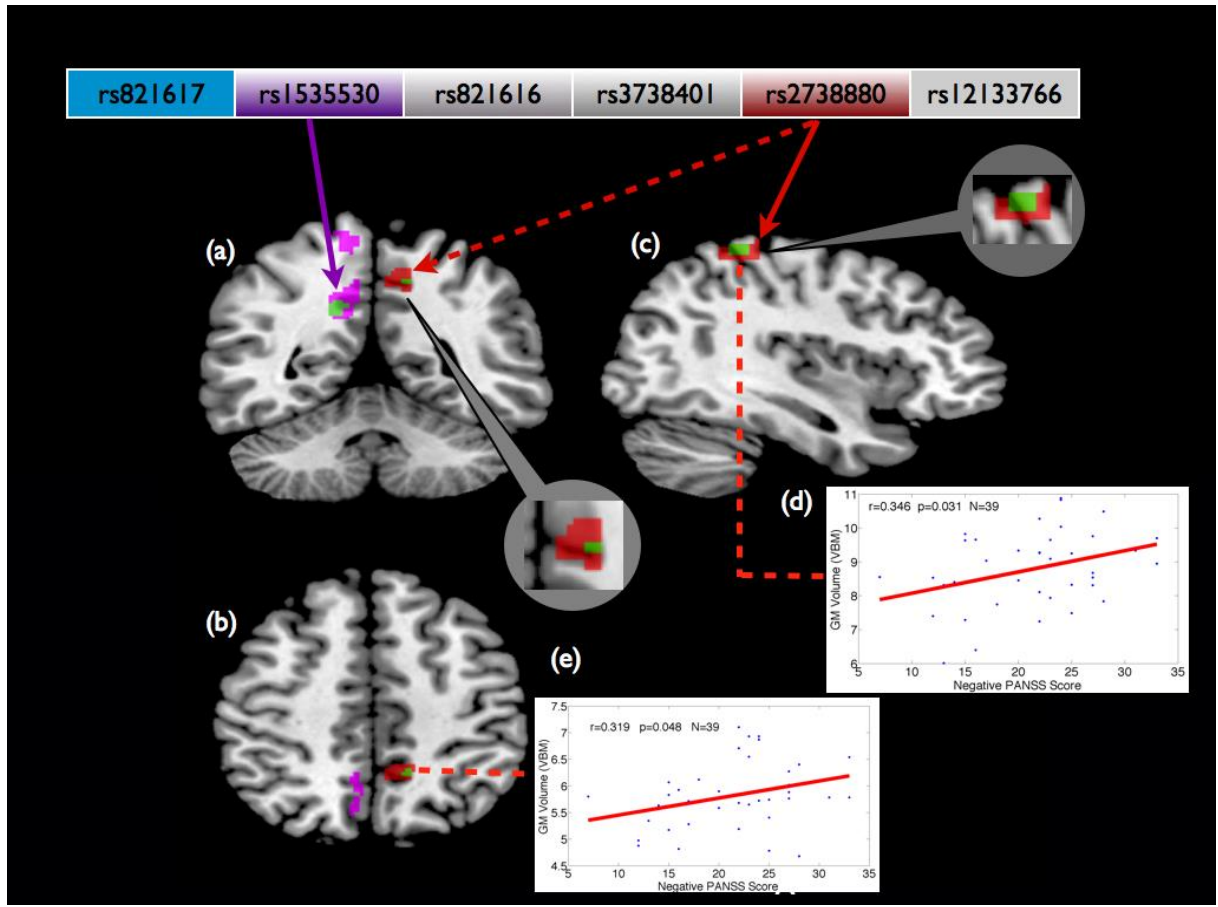

**Figure 3.** Location of significant DISC1 associated GM from the VWAS analysis and correlation with PANSS scores. Sections (a) and (b): The purple (left) and red (right) regions show significant clusters in the right and left precuneus respectively, identified by the cluster analysis with six *DISC1* SNPs; The inset green sub-regions in the purple and red regions show the significant voxels in the corresponding right and left precuneus respectively, identified by the peak analysis with DISC SNPs, rs1535530 and rs2738880, indicated by the purple and red arrows respectively (the dash red arrow shows that relation is not significant but trends to be significant). Section (c): the red region shows significant

cluster in the right post-central gyrus, identified by the cluster analysis with six *DISC1* SNPs; the inset green sub-region in the red region shows the significant voxels in the corresponding right post-cntral gyrus, identified by the peak analysis with a single DISC SNP rs2738880, indicated by the red arrow. The inset regression plots (d and e) show the significant correlations between the GM volumes of significant voxels in the right post-central gyrus (d) and left precuneus (e), and the PANSS negative scores.

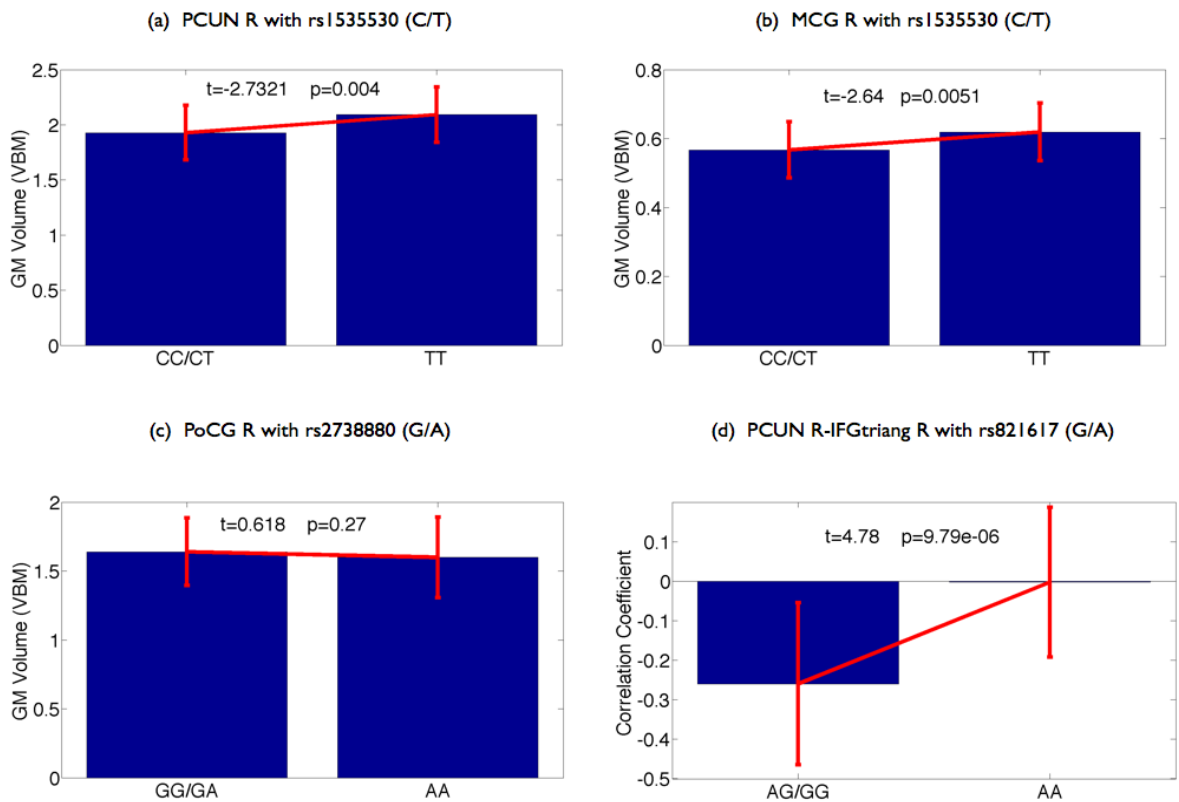

**Figure 4.** The comparison between the GM volumes of the significant voxels identified by VWAS and the correlation coefficients of the links identified by LWAS for different genotype carriers: (a). GM of PCUN R with rs1535530; (d). GM of MCG R with rs1535530; (c). GM of PoCG R with rs2738880; (d) Correlation coefficients of PCUN R-IFGtriang R with rs821617. Here, blue bars stands for SE and red for SD.

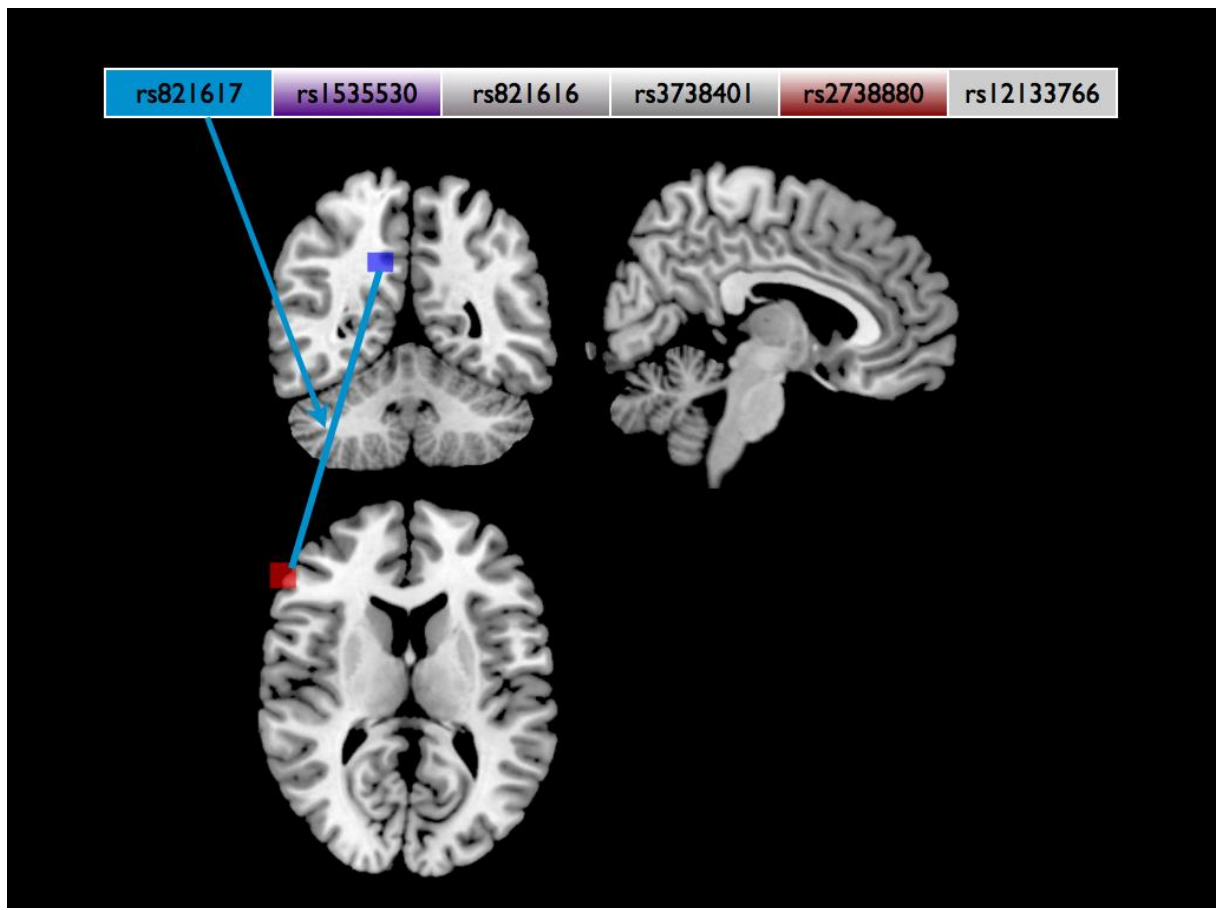

**Figure 5.** Location of significant DISC1 associated functional connections from the LWAS analysis. The blue region shows the sphere seed ROI in the right precuneus; the red region shows the associated target cube ROIs, located in the triangular inferior frontal gyrus identified with the six common SNPs in DISC1, as well as its significantly correlated individual SNP rs821617 respectively.

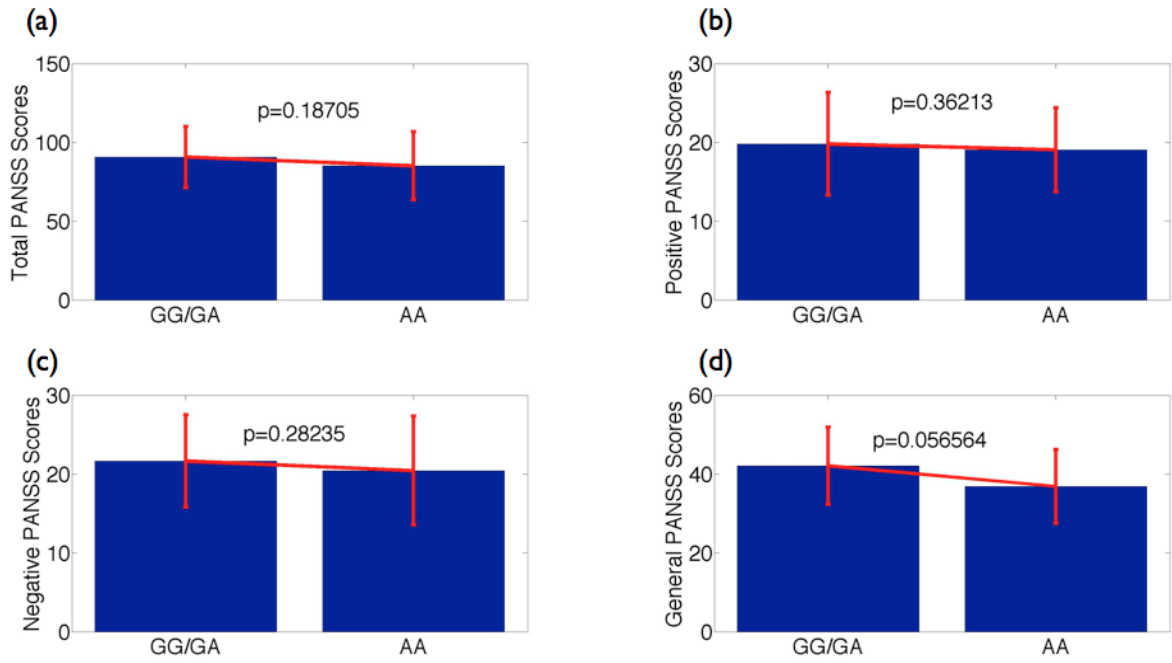

**Figure 6.** Comparison of total (a), positive (b), negative (c) and general (d) PANSS scores between the genotypes GG/GA and the genotype AA in the SNP rs2738880, where the p-values are derived by two-sample student t-test, with blue bars for SE and red for SD.

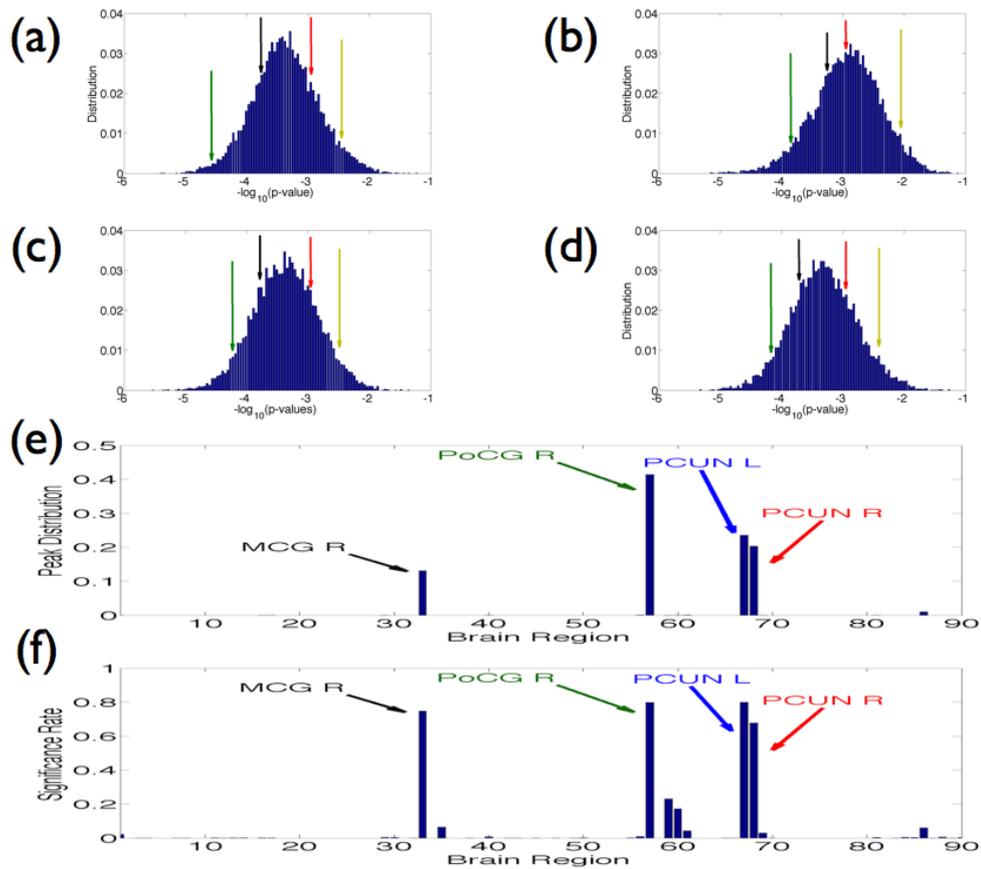

**Figure 7.** Empirical distributions of chi-square statistic p-values for the peak voxels in the right precuneus (PCUN) (a), left precuneus (b), right post-central gyrus (PoCG) (c) and right middle cingulate gyrus (MCG) (d), where the green and yellow arrows show the 5% and 95% limits of the distribution, and the black and red the original p-values and threshold (AAL brain region-wise Bonferroni correction); the distribution of the location of peak voxels over all GM with respect to 90 AAL brain regions (e); percents of each of the 90 AAL brain regions containing at least one significant voxel in 10000 times of random resampling (f).

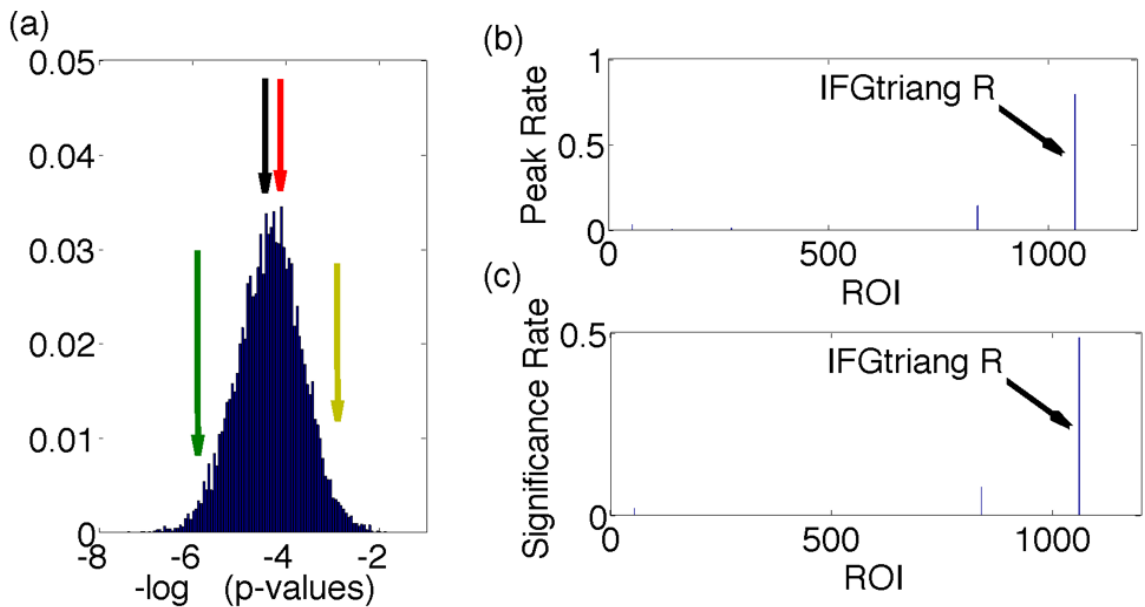

**Figure 8.** Empirical distributions of the chi-square statistics p-values for the significant link between ROIs in the right precuneus (seed) and right triangular inferior frontal gyrus (a), where the green and yellow arrows stand indicate 5% and 95% limits of the distribution, and the black and red the original p-values and threshold (1072 cube ROI-wise Bonfferroni correction); the distribution of the location of the cube ROI of the peak link over all 1072 cube brain ROIs from the seeds in the right precuneus (b); percents of each 1072 cube ROIs containing at least one significant link in 10000 times of random resampling from the seeds in the right precuneus (c).

**Table 1.** Subject Demographics.

| Group | No. of Subjects | Sex (M/F) | Age (years) | Illness Duration (years) | Education (years) | PANSS[2] Scores | | | |
|---|---|---|---|---|---|---|---|---|---|
| | | | | | | Total | Positive | Negative | General |
| Schizophrenia patients | 46(8 UM[1]) | 27/19 | 24.2±5.7 | 18.1±15.9 | 12.6±2.4 | 88.8±20.1 | 19.5±6.1 | 21.2±6.2 | 40.2±9.8 |
| Healthy controls | 24 | 14/10 | 28.3±5.9 | N/A | 13.2±3.6 | N/A | N/A | N/A | N/A |
| p values | N/A | 0.9165[3] | 0.0061[3] | N/A | 0.375 | N/A | N/A | N/A | N/A |

*[1]"UM" means the number of the unmediated subjects in the schizophrenia group;*

*[2] Among 46 schizophrenia patients, there are 7 subjects who have no PANSS component scores only total scores.[3] t-test with df=68.*

**Table 2.** VWAS Results.

| Peak Analyses | | | | | | | | |
|---|---|---|---|---|---|---|---|---|
| SNPs | Regions | Talairach coordinates (mm) of the peak voxel in the region | Chi-square statistics and p-values of the peak voxel (corrected) | No. of significant voxels | Grey Matter Volume and PANSS correlation scores (p-values) | | | |
| | | | | | Total (N=46) | Positive (N=39) | Negative (N=39) | General (N=39) |
| Six common SNPs[1] | PCUN R | [14,-52,36] | $\chi^2$=**26.63**, **p=1.694e-4 (0.0076)**[2] | 13 | 0.002 (0.990) | -0.069 (0.676) | 0.032 (0.845) | -0.067 (0.686) |
| | PCUN L | [-10,-64,64] | $\chi^2$=**23.852**, **p=5.562e-4 (0.025)** | 20 | 0.103 (0.494) | -0.138 (0.401) | **0.319 (0.048)** | 0.080 (0.631) |
| | MCG R | [14,-50,34] | $\chi^2$=**26.319**, **p=1.942e-4 (0.0087)** | 9 | -0.034 (0.824) | -0.046 (0.782) | -0.044 (0.790) | -0.084 (0.613) |
| | PoCG R | [38,-40,64] | $\chi^2$=**26.716**, **p=1.636e-4 (0.0074)** | 29 | **0.352 (0.017)** | 0.180 (0.274) | **0.346 (0.031)** | 0.312 (0.053) |
| rs2738880[3] | PCUN L | [-12,-48,50] | $\chi^2$=**16.731**, **p=2.328-4 (0.063)** | 0 | *N/A* | N/A | *N/A* | N/A |
| | PoCG R | [40,-40,64] | $\chi^2$=**18.353**, **p=1.034e-5 (0.0279)** | 6 | **0.338 (0.022)** | 0.168 (0.307) | **0.342 (0.033)** | 0.302 (0.062) |
| rs1535530[1] | PCUN R | [14,-52,36] | $\chi^2$=**16.066**, **p=6.1175e-5 (0.0165)** | 6 | -0.033 (0.827) | -0.090 (0.587) | -0.014 (0.931) | -0.093 (0.575) |
| | MCG R | [14,-50,34] | $\chi^2$=**15.499**, **p=8.254e- 5(0.0223)** | 2 | -0.050 (0.743) | -0.048 (0.774) | -0.059 (0.722) | -0.088 (0.596) |
| Cluster Analyses The Largest clusters (in RESELS) and their centre regions | | | | | | | | |
| SNPs | No. | RESELs/Voxels/Mean Volumes (VBM) | P-values (corrected) | Centre Regions | | | | |
| Six common SNPs | 1 | 7. 2925/584/144.01 | ***2.039e-05 (1.572e-04)*** | PCUN R, MCG R | | | | |
| | 2 | 1.5373/102/25.53 | 0.013 (0.096) | PCUN L | | | | |
| | 3 | 1.4617/146/32.42 | 0.015 (0.108) | PoCG R, SPG R | | | | |
| | 4 | 1.334/134/28.21 | 0.019 (0.135) | PCUN L | | | | |
| rs2738880 | 1 | 3.404/347/78.49 | **0.0018 (0.0114)** | PoCG R, SPG R | | | | |
| | 2 | 3.181/360/89.45 | **0.0023 (0.0147)** | PCUN R, MCG R | | | | |
| | 3 | 2.5694/200/52.92 | **0.0047 (0.0302)** | PCUN L | | | | |
| rs1535530 | 1 | 5.337/250/63.81 | **4.93e-4 (0.0028)** | PCUN R, MCG R | | | | |
| | 2 | 1.6021/60/15.65 | 0.025 (0.132) | MTG L | | | | |
| | 3 | 1.422/85/18.77 | 0.033 (0.169) | MFG R | | | | |

[1] $df=6$ for chi-square statistics; [2] Bold font means that the corrected p-value is less than 0.05;

[3, 4] $df=1$ for chi-square statistics; [5] Bold and italic font means that the corrected p-value is less than 0.001.

Table 3. LWAS Results

| SNPs | Regions of Link ends | Talairach coordinates (mm) of Seed Centres | Talairach coordinates (mm) of Target centres | Chi-square statistics and P-values (corrected) | Correlation Coefficients and PANSS correlation scores (p-values) | | | |
|---|---|---|---|---|---|---|---|---|
| | | | | | Total (N=46) | Positive (N=39) | Negative (N=39) | General (N=39) |
| Six common SNPs[1] | PCUN R-IFGtriang R | [14,-52,36] | [57,36,6] | **$\chi^2$=32.2929, p=1.434e-05 (0.0154)**[2] | 0.111 (0.461) | 0.160 (0.329) | -0.0031 (0.985) | 0.125 (0.450) |
| | PCUN L-ORBmid R | [-10,-64,64] | [38,41,-18] | $\chi^2$=27.278, p=1.284e-04 (0.138) | -0.029 (0.851) | -0.018 (0.925) | -0.082 (0.620) | -0.106 (0.522) |
| | PoCG R-MFG R | [38,-40,64] | [34,38,32] | $\chi^2$=27.202, p=1.327e-4 (0.142) | 0.090 (0.548) | -0.013 (0.928) | 0.196 (0.232) | 0.103 (0.530) |
| | MCG R-IFGtriang R | [14,-50,34] | [18,-20,2] | $\chi^2$=28.4993, p=7.565e-05 (0.081) | 0.065 (0.667) | 0.141 (0.393) | 0.026 (0.873) | 0.060 (0.716) |
| rs821617[3] | PCUN R-IFGtriang R | [14,-52,36] | [57,36,6] | **$\chi^2$=20.697, p=5.381e-6 (0.0346)** | N/A | N/A | N/A | N/A |
| rs12133766[4] | PCUN L-PCUN L | [-14,-50,48] | [-10,-50,48] | $\chi^2$=15.508, p=4.2911e-04 (>1) | N/A | N/A | N/A | N/A |

[1] *df=6 chi-square statistics;* [2] *Bold font means the corrected p-values are less than 0.05;* [3] *df=1 for chi-square statistics.*